\documentclass[aps,prb,reprint,showkeys]{revtex4-1}
\usepackage{bm,graphicx,tabularx,array,dcolumn,xcolor,microtype,multirow,amscd,amsmath,amssymb,amsfonts,physics}
\usepackage[version=4]{mhchem}
\usepackage[utf8]{inputenc}
\usepackage[T1]{fontenc}
\usepackage{txfonts}

\usepackage{dblfloatfix,placeins,float,physics}

\usepackage[
	colorlinks=true,
    citecolor=blue,
    linkcolor=blue,
    filecolor=blue,      
    urlcolor=blue,	
    breaklinks=true
	]{hyperref}
\usepackage{caption} % Allow width of caption to be controlled
\usepackage{subcaption}

\newcommand{\CAM}{Yusuf Hamied Department of Chemistry, Lensfield Rd, Cambridge CB2 1EW}

\begin{document}

\title{Enhanced sampling reveals the main pathway of an organic cascade reaction}
\author{Yi Sun}
\email{ys595@cam.ac.uk}
\affiliation{\CAM}

\author{Xu Han, Lijiang Yang}
\email{yanglj@pku.edu.cn}
\affiliation{Beijing National Laboratory for Molecular Sciences, College of Chemistry and Molecular Engineering, Peking University, Beijing 100871, China;  Institute of Systems and Physical Biology, Shenzhen Bay Laboratory, Shenzhen, Guangdong 518132, China}

\date{\today}

\begin{abstract}
Chemical reactions are usually hard to be simulated using conventional molecular simulations since the transition state is difficult to be captured. Therefore, enhanced sampling methods are implemented to accelerate the occurrence of chemical reactions. In this investigation, we present an application of metadynamics in simulating an organic multi-step cascade reaction. The analysis of the reaction trajectory reveals the barrier heights of both forward and reverse reactions. We also present a discussion of the advantages and disadvantages of generating the reactive pathway using enhanced sampling molecular simulations and the intrinsic reaction coordinate (IRC) algorithm.
\end{abstract}

\maketitle
\raggedbottom

%%%%%%%%%%%%%%%%%%%%%%%%%%%%%%%%%%%%%%%%%%%%%%%%%%%%%%%%%%%%%%
\section{Introduction}
%%%%%%%%%%%%%%%%%%%%%%%%%%%%%%%%%%%%%%%%%%%%%%%%%%%%%%%%%%%%%%
Molecular simulations are gaining importance in physics, chemistry, biology, and materials research. Due to the high computational costs, it is difficult to investigate a range of natural phenomena requiring rare events, such as those exhibited in phase transitions, chemical processes, and protein folding, using traditional molecular dynamics (MD). Individual states in these systems are separated by colossal free energy barriers. Thus, the transition between them takes aeons.

Performing enhanced sampling simulations, which are often divided into two categories: collective variable (CV)-based and CV-free approaches, is a solution to this issue. CVs characterise the most challenging modes to sample and are typically used to distinguish between metastable states. In order to accelerate the transition between metastable states, CV-based approaches such as umbrella sampling (US) \cite{Torrie} and metadynamics (MetaD) \cite{Laio} can improve sampling over CVs. CV-free methods, such as replica exchange MD (REMD) \cite{Sugita} and integrated tempering sampling (ITS) \cite{Gao}, can facilitate transitions between distinct metastable states with little a priori system knowledge. Recently, hybrid approaches, which mix the two categories of methods \cite{Bussi,Yang} , can further improve the sampling capabilities over the required configuration or phase space \cite{Yang2,Yang3}.

There are already studies that apply enhanced sampling simulations in organic reactions\cite{Zhang,Nottoli,Han}. However, most of them only concern single-step reactions (although a pair of stereoisomers can be formed). In this study, a metadynamics simulation is implemented in a two-step reaction that includes two different types of pericyclic reactions and has significance in synthetic chemistry\cite{Nico,Paquette}. The scheme of which is shown in Figure 1. 

\begin{figure}[H]
\includegraphics[width=\linewidth]{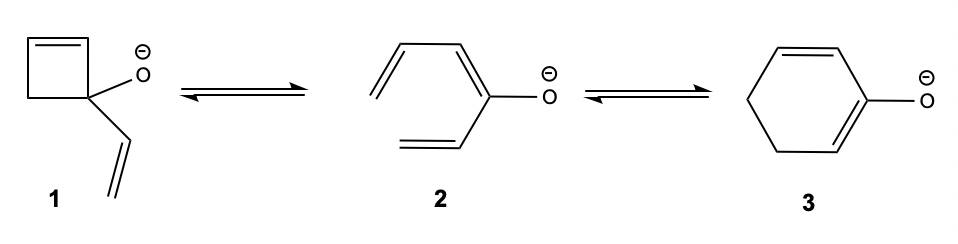}
\caption{A double-step organic cascade reaction}
\centering
\end{figure}

We tried to figure out the reaction pathway and obtain the free energy barrier height through the metadynamics simultaion. A discussion is followed to evaluate the efficiency of finding the pathway using enhanced sampling and IRC. 

%%%%%%%%%%%%%%%%%%%%%%%%%%%%%%%%%%%%%%%%%%%%%%%%%%%%%%%%%%%%%%
\section{Theory}
%%%%%%%%%%%%%%%%%%%%%%%%%%%%%%%%%%%%%%%%%%%%%%%%%%%%%%%%%%%%%%

\emph{Metadynamics (MetaD)}: The general idea behind MetaD is to prevents an ergodic sampling by adding a Gaussian biased potential $V_{bias}$ to the system Hamiltonian\cite{Laio,Yang4}. The biased potential is a function of collective variables (CVs) $\textbf{s}$. A collective variable is a function of the particle positions $\textbf{r}$.

\begin{equation} \label{eq:1}
V_{bias}(\textbf{s},t)=\int_{0}^{t} G\left[\textbf{s},\textbf{s}(\tau)\right] d\tau
\end{equation}

\begin{equation} \label{eq:2}
G\left(\textbf{s},\textbf{s}^{'}\right)=\omega \cdot \text{exp}\left[-\frac{\left(\textbf{s}-\textbf{s}^{'}\right)^2}{2\boldsymbol{\sigma}^2}\right]
\end{equation}

where $t$ is the simulation time. $\omega$ is the height of the Gaussian, and $\boldsymbol{\sigma}$ is the standard deviation vector. 

The goal of a MetaD run is to achieve the convergence of the free energy landscape. Theoretically speaking, at the end of a MetaD run, the value of the free energy $F(\textbf{s})$ is equal to the negative value of the accumulated bias potential, i.e., 

\begin{equation} \label{eq:3}
F(\textbf{s})= -\lim_{t \rightarrow \infty}V(\textbf{s},t)
\end{equation}

However, the constant addition of the repulsive potential actually prevents the convergence of the free energy surface, introducing a systematic error into it \cite{Laio2,Bussi2,Barducci}. The way to tackle this problem is to introduce a new method known as the well-tempered metadynamics (WT-MetaD), which is used in this study \cite{Barducci2}. 

\emph{Well-tempered metadynamics (WT-MetaD)}: The biggest difference between MetaD and WT-MetaD is that the height of the gaussian $\omega$ is now time-dependent. 

\begin{equation} \label{eq:4}
\omega(t)=w \cdot\text{exp}\left(-\frac{1}{\gamma-1}\beta V_{bias}\right)
\end{equation}

where $w$ is the height of the first gaussian, $\gamma>1$ is known as the bias factor. As the height of the gaussian $\omega(t)$ decreases as the bias potential increases. When $t \rightarrow \infty$, $\omega(t) \rightarrow 0$, and the bias potential converges as 

\begin{equation} \label{eq:5}
\lim_{t \rightarrow \infty}V(\textbf{s},t) = -\left(1-\frac{1}{\gamma}\right) F(\textbf{s})
\end{equation}

The free energy landscape $F(\textbf{s})$ can then be calculated. 

%%%%%%%%%%%%%%%%%%%%%%%%%%%%%%%%%%%%%%%%%%%%%%%%%%%%%%%%%%%%%%
\section{System and method}
%%%%%%%%%%%%%%%%%%%%%%%%%%%%%%%%%%%%%%%%%%%%%%%%%%%%%%%%%%%%%%

The simulated reaction is shown in Figure 2. The QM/MM based molecular dynamics simulation is performed on SANDER in AMBER18 \cite{Case}, and the enhanced sampling plugin to perform WT-MetaD is the open-source, community-developed PLUMED library \cite{PLUMED}, version 2.5.1 \cite{PLUMED2}. In order to fit the experimental conditions, the simulation is in the NVT emsemble, and the temperature is set to be 200K, and a Langevin thermostat with a friction coefficient of 1 $ps^{-1}$ is enabled to control the temperature of the system. An implicit solvation model is used to model the solvent (THF). The conditions are chosen to fit the experimental conditions \cite{Nico,Paquette}. 

The simulation length is 20 ns with four parallel trajectories. As there are several millions of configurations, the level of energy calculation is set to be DFTB3 in order to reduce time cost \cite{Porezag, Seabra}. The accuracy of the DFTB method for barrier heights of a variety of reactions has been verified \cite{Gruden}. The two CVs are selected to be the two bonds that is broken and formed during the whole reaction, also shown in Figure 2. 

\begin{figure}[H]
\includegraphics[width=\linewidth]{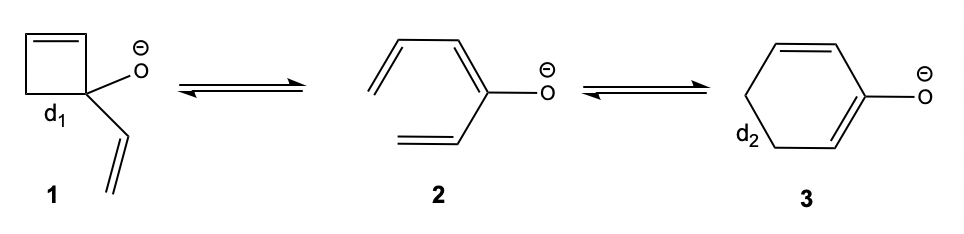}
\caption{The simulated reactions and the two CVs: $d_1$ and $d_2$.}
\centering
\end{figure}

The reaction pathway is then sketched using the MEPplot package \cite{MEP}. The initial points are placed at the bottom of the free energy landscape of each compound. The free energy pathway is then elucidated after its convergence reported from the package.

%%%%%%%%%%%%%%%%%%%%%%%%%%%%%%%%%%%%%%%%%%%%%%%%%%%%%%%%%%%%%%
\section{Results and Discussion}
%%%%%%%%%%%%%%%%%%%%%%%%%%%%%%%%%%%%%%%%%%%%%%%%%%%%%%%%%%%%%%

The effectiveness of MetaD simulations is examined by plotting the change of the magnitude of CVs versus time. The relationship between $d_1$ and $t$, and that between $d_2$ and $t$ in a trajectory is shown in Figure 3 and 4, respectively. For the clarity of the plot, one point is printed per 1000 fs. 
\begin{figure}[H]
\includegraphics[width=\linewidth]{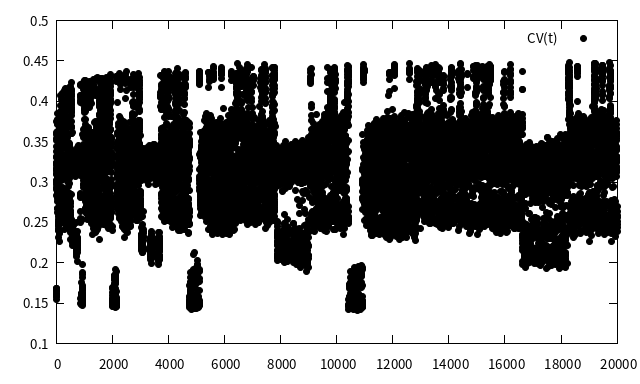}
\caption{The variation of $d_1$ through time. The horizontal axis represents the simulation time in ps, and the vertical axis represents the length of $d_1$ in nm. }
\centering
\end{figure}
\begin{figure}[H]
\includegraphics[width=\linewidth]{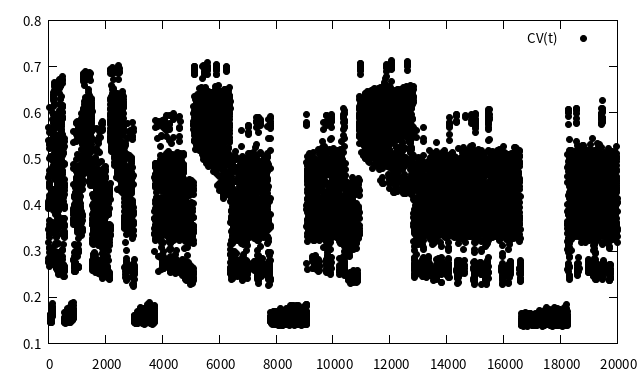}
\caption{The variation of $d_2$ through time. The horizontal axis represents the simulation time in ps, and the vertical axis represents the length of $d_2$ in nm. }
\centering
\end{figure}

According to the graphs, both CVs have changed considerably through time, which implies multiple conversion between compounds. For the case of $d_1$, a distance around 0.15 nm corresponds to compound 1. All the other distances corresponds to multiple isomers of the acyclic compound 2 or compound 3. For the case of $d_2$, a distance around 0.15 nm corresponds to compound 3, All the other distances corresponds to multiple isomers of the acyclic compound 2 or compound 1.

The part of free energy landscape of this trajectory with the reaction trajectory is sketched, shown in Figure 5. The free energy curve is shown in Figure 6. 
\begin{figure}[H]
\includegraphics[width=\linewidth]{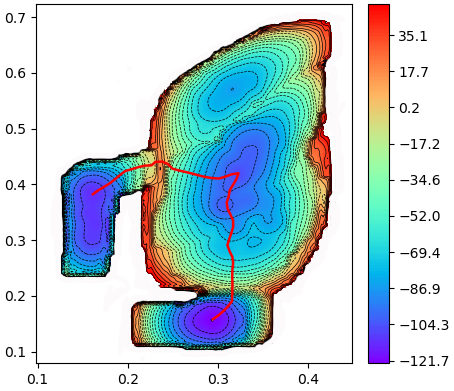}
\caption{The free energy surface and the trajectory. The horizontal axis corresponds to $d_1$ in nm. The vertical axis corresponds to $d_2$ in nm. The colours in the graph depict the energy value at a certain combination of distances in $kJ/mol$.}
\centering
\end{figure}
\begin{figure}[H]
\includegraphics[width=\linewidth]{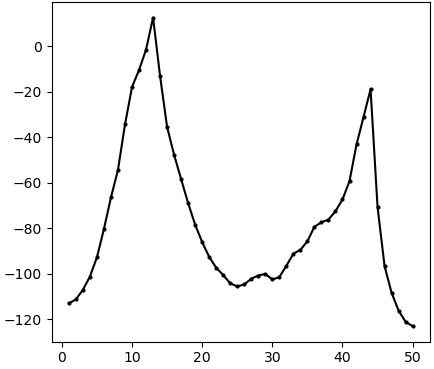}
\caption{The potential energy curve. The dimensionless horizontal axis corresponds to the reaction coordinate and the numbers represent the progression. The vertical axis corresponds to free energy, whose unit is in $kJ/mol$.}
\centering
\end{figure}
The barrier heights of the forward and the reverse reactions are also measured, shown in the table below:
\begin{table}[H]
\begin{tabular}{|c|c|c|c|c|}
\hline
 & 1 $\rightarrow$ 2 & 2 $\rightarrow$ 1 & 2 $\rightarrow$ 3 & 3 $\rightarrow$ 2 \\ \hline
Barrier Height (kcal/mol) & 28.7 (1.5)      & 25.7 (1.9)      & 21.2 (0.4)      & 26.4 (1.2)      \\ \hline
\end{tabular}
\caption{The barrier heights. The number in the brackets are the error bars.}
\label{tab:table1}
\end{table}

As the objective of this study is to generate the reaction pathway via enhanced sampling methods. This is a relatively novel approach, as the traditional way of doing this is to perform the IRC algorithm\cite{Fukui, Hratchian}. It is worth while to discuss the differences between them, and their strength and weaknesses. 

IRC algorithms require the starting compound to be the transition state if a complete trajectory is required. This is because, in principle, the IRC-LQA algorithm operates as follows \cite{Page, Page2}:

The starting molecule, which has a position vector $\textbf{x}_{\textbf{0}}$, can expand its potential energy surface (PES) around it. The energy is expressed as 
\begin{equation} \label{eq:6}
E(\textbf{x})=E_0+\textbf{g}_\textbf{0}^\textbf{t} \Delta\textbf{x}+\frac{1}{2}\Delta\textbf{x}^\textbf{t}\textbf{H}_\textbf{0}\Delta\textbf{x}
\end{equation}
where $\Delta \textbf{x}$ is the displacement vector, $\textbf{g}_\textbf{0}$ is the energy gradient at $\textbf{x}_{\textbf{0}}$, and $\textbf{H}_\textbf{0}$ is the Hessian matrix at $\textbf{x}_{\textbf{0}}$. Taking the first derivative of equation (\ref{eq:6}) with respect to $\textbf{x}$ give the energy gradient at $\textbf{x}$ as
\begin{equation} \label{eq:7}
\textbf{g}_\textbf{0}\textbf{(x)}= \textbf{g}_\textbf{0}+\textbf{H}_\textbf{0}\Delta \textbf{x}
\end{equation}

Then the coordinate vector $\textbf{x}$ can be updated using the steepest descent equation
\begin{equation} \label{eq:8}
\frac{\text{d}\textbf{x}(s)}{\text{d}s}=-\frac{\textbf{g}(\textbf{x})}{\abs{\textbf{g}(\textbf{x})}}=-\frac{\textbf{g}_\textbf{0}+\textbf{H}_\textbf{0}\Delta \textbf{x}}{\abs{ \textbf{g}_\textbf{0}+\textbf{H}_\textbf{0}\Delta \textbf{x}}}
\end{equation}
where $s$ is an arc along the reaction path.

The most important prerequisite of a successful IRC calculation is that the input structure must be the transition state (TS). This is because the IRC calculation is usually calculated on both sides, i.e., the position vector is updates on opposite directions. If the starting compound is not a TS, the updated energy value would rise, causing the IRC algorithm to halt. There are two problems in performing IRC calculations.

The first problem in IRC is that finding the TS is far more non-trivial than finding the energy minima. In practise, a number of failures are always accompanied with finding the TS, especially when the implicit solvation model is added. As a TS is essential in performing an IRC calculation, being unable to find it will significantly hamper the progress to find the reaction path.

The second problem is that IRC, just like other TS-related algorithms, requires the Hessian matrix of a molecule \cite{Page}. As a Hessian matrix is found by evaluating the second derivative of the molecular energy with respect to two atomic movements, the number of energy calculations is proportional to $N^2$, where N is the number of atoms in the molecule. This makes IRC calculations very computational expensive at high levels of theory. 

On the contrary, molecular dynamics simulations with enhanced sampling are more straight forward to carry out. In theory, one only need to provide a structure of molecule, which does not even need to be at the energy minima, and (a) reaction coordinate(s) as the CV. After some time period (usually 20 to 30 ns), the FEL landscape will converge, and the minimum energy pathway can be found by using made packages, e.g. MULE \cite{MULE} or MEP Plot \cite{MEP}. The benefit of a completed FEL, compared to a single path obtained via IRC is that it may potentially reveal more reactive pathways than the most feasible one. However, obtaining the converged FEL is easier said than done.

One of the main problems of enhanced sampling is that the molecule would collapse after adding a bias potential. Even if the bond is not set as a CV, the bias added will still significantly affect the bond. As a result, it may break after sometime, making the whole simulation unable to proceed. The way of countering this is to add upper walls to the molecule to prevent the bonds from breaking. Such an issue is particularly serious in this study due to the high angle strain of the four-membered ring. Consequently, apart from the two bonds that are going to form(break) during the reaction, upper walls are added to all other carbon-carbon bonds and carbon-oxygen bonds. 

Another issue of enhanced sampling is that there may be expected products. This, again, is due to the bias added to the system so that it is chemically labile. In this study, an attempt was made to simulate the following reaction, shown in Figure 7 with CVs shown in the picture. 
\begin{figure}[H]
\includegraphics[width=\linewidth]{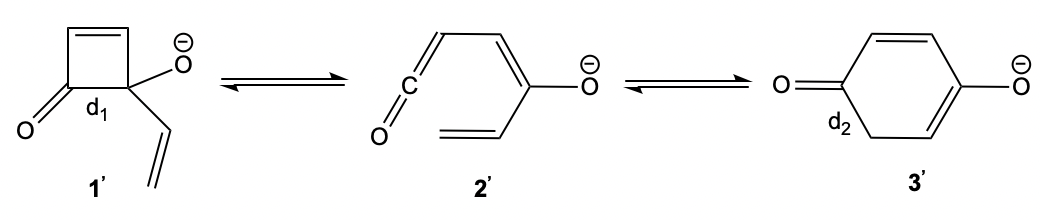}
\caption{The reaction that was attempted to simulate}
\centering
\end{figure}

The main problem of performing an enhanced sampling simulation to this reaction is the conversion from compound 2' to compound 6', shown in Figures 8 and 9 below
\begin{figure}[H]
\includegraphics[width=\linewidth]{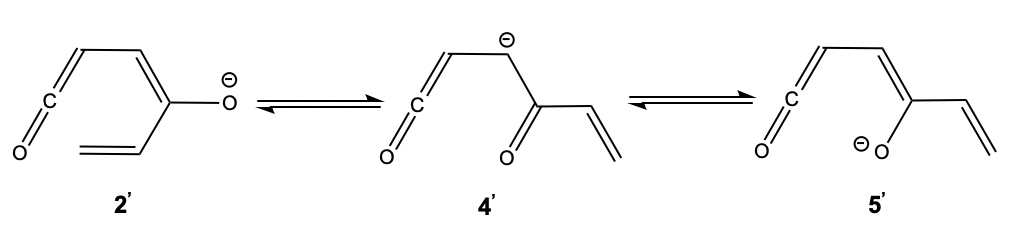}
\caption{The isomerisation from compound 2' to 5'.}
\centering
\end{figure}
\begin{figure}[H]
\includegraphics[width=\linewidth]{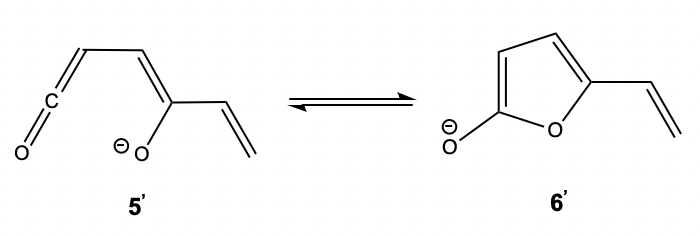}
\caption{The addition to the chemically active ketene.}
\centering
\end{figure}

Comparing to compound 5', compound 6' not only does not contain any highly unsaturated ketene structures, but also has an aromatic furan ring. The high stability of compound 6 essentially prevents the simulation from proceeding. In practise, the FEL does not converge even after 200 ns of simulation, even when a lower wall is added to partially stop the formation of the expected carbon-oxygen bond. In general, the number of unexpected byproducts will increase significantly when the system gets larger, which creates huge difficulty in simulating them.

Another problem lies in the theory of calculation, semi-empirical and force field based energy algorithms are the only feasible ones to perform energy calculations of a colossal number of configurations (one million per nanosecond). Any attempts to refine the precision of FEL, even using 6-31(G) level of theory make takes days to complete even under a supercomputer.

The final issue is the efficiency of enhanced sampling. Even though the transition between compounds is possible due to the addition of a biased potential, most of the simulation time is still wasted. This issue is also profound in this study, as the acyclic compound 2 has a large number of stereoisomers. Consequently, even though the simulation lasts for 20ns, only 20 transitions between the reacatants are observed, and it takes progressively longer for a new transition to occur. 

%%%%%%%%%%%%%%%%%%%%%%%%%%%%%%%%%%%%%%%%%%%%%%%%%%%%%%%%%%%%%%
\section{Conclusion and future work}
%%%%%%%%%%%%%%%%%%%%%%%%%%%%%%%%%%%%%%%%%%%%%%%%%%%%%%%%%%%%%%
In this study, an enhanced sampling simulation is performed on an organic cascade reaction. The FEL is then sketched as well as the potential energy curve and the barrier heights. A comparison is then drawn to compare the feasibility of finding the reaction pathway via IRC and enhanced sampling.

Based on the current deficiencies of enhanced sampling, three future directions are suggested:

(1) Add the option in the molecular dynamics package to prevent irrelevant bond from breaking and unneeded side products.

(2) Designing algorithms that can improve the efficiency of the enhanced sampling simulations, especially when sampling the TS.

(3) Apply enhanced sampling molecular simulations in more organic cascade reactions to further demonstrate the benefits of this method. 

%%%%%%%%%%%%%%%%%%%%%%%%%%%%%%%%%%%%%%%%%%%%%%%%%%%%%%%%%%%%%%
\section*{Acknowledgements}
%%%%%%%%%%%%%%%%%%%%%%%%%%%%%%%%%%%%%%%%%%%%%%%%%%%%%%%%%%%%%%
The authors thank Yi Yang, Yiqin Gao and Sizhe Li for providing meaningful discussions.

%%%%%%%%%%%%%%%%%%%%%%%%%%%%%%%%%%%%%%%%%%%%%%%%%%%%%%%%%%%%%%
%\section*{References}

%%%%%%%%%%%%%%%%%%%%%%%%%%%%%%%%%%%%%%%%%%%%%%%%%%%%%%%%%%%%%%

\end{document}